%% file: ms.tex
\documentclass[conference]{IEEEtran}
\IEEEoverridecommandlockouts
\usepackage{cite}
\usepackage{amsmath,amssymb,amsfonts}
\usepackage{algorithmic}
\usepackage{graphicx}
\usepackage{textcomp}
\usepackage{xspace}  
\usepackage{tabularx}
\usepackage{xcolor}  
\usepackage{multirow}   
\usepackage{booktabs}   
\usepackage{hyperref}  
\usepackage{makecell} 
\usepackage{pgfplots}   
\usepackage{flushend}
\usepackage{balance}
\usepackage{listings}
\usepackage{float} 
  
\definecolor{codegreen}{rgb}{0,0.6,0} 
\definecolor{codegray}{rgb}{0.5,0.5,0.5}
\definecolor{codepurple}{rgb}{0.58,0,0.82} 
\definecolor{backcolour}{rgb}{1,1,1}
\lstdefinestyle{codeStyle}{
    backgroundcolor=\color{backcolour},   
    commentstyle=\color{codegreen},
    keywordstyle=\color{magenta},
    numberstyle=\tiny\color{codegray},
    stringstyle=\color{codepurple},
    basicstyle=\ttfamily\footnotesize,
    breakatwhitespace=false,         
    breaklines=true,                 
    captionpos=b,                    
    keepspaces=true,                 
    numbers=none,                    
    numbersep=5pt,                  
    showspaces=false,                
    showstringspaces=false,
    showtabs=false,  
    frame=single,
    tabsize=2,
}

\lstdefinestyle{promptStyle}{
    backgroundcolor=\color{backcolour},   
    commentstyle=\color{codegreen},
    numberstyle=\tiny\color{codegray},
    basicstyle=\sffamily\scriptsize,
    breakatwhitespace=false,         
    breaklines=true,                 
    captionpos=b,                    
    keepspaces=true,                 
    numbers=none,                    
    showspaces=false,                
    showstringspaces=false,
    showtabs=false,                  
    frame=single,
    tabsize=2
}

\newcommand{\rev}[1]{\textcolor{black}{#1}}

\newboolean{showcomments}
\setboolean{showcomments}{true} % toggle to show or hide comments
\ifthenelse{\boolean{showcomments}} 
{
  \newcommand{\nb}[2]{  
    \fcolorbox{black}{yellow}{\bfseries\sffamily\scriptsize#1} 
    {\sf\small$\blacktriangleright$\textit{#2}$\blacktriangleleft$}
  } 
} 
{
  \newcommand{\nb}[2]{} 
}

\newcommand*\circled[1]{\tikz[baseline=(myanchor.base)] \node[circle,fill=.,inner sep=1pt] (myanchor) {\color{-.}\bfseries\footnotesize #1};}

\begin{document}

\title{A Bot-based Approach to Manage Codes of Conduct in Open-Source Projects}

\author{
\IEEEauthorblockN{Sergio Cobos}
\IEEEauthorblockA{Universitat Oberta de Catalunya\\
Barcelona, Spain \\
scobosga@uoc.edu}
\and
\IEEEauthorblockN{Javier Luis Cánovas Izquierdo}
\IEEEauthorblockA{Universitat Oberta de Catalunya\\
Barcelona, Spain \\
jcanovasi@uoc.edu}
}

\maketitle

\begin{abstract}
The development of Open-Source Software (OSS) projects relies on the collaborative work of contributors, generally scattered around the world.
To enable this collaboration, OSS projects are hosted on social-coding platforms like GitHub, which provide the infrastructure to host the code as well as the support for enabling the participation of the community.
The potentially rich and diverse mixture of contributors in OSS projects makes their management not only a technical challenge, where automation tools and bots are usually deployed, but also a social one.
To this aim, OSS projects have been increasingly deploying a declaration of their code of conduct, which defines rules to ensure a respectful and inclusive participatory environment in the community, being the Contributor Covenant the main model to follow.
However, the broad adoption and enforcement of codes of conduct in OSS projects is still limited.
In particular, the definition, deployment, and enforcement of codes of conduct is a very challenging task.
In this paper, we propose an approach to effectively manage codes of conduct in OSS projects based on the Contributor Covenant proposal.
Our solution has been implemented as a bot-based solution where bots help in the definition of codes of conduct, the monitoring of OSS projects, and the enforcement of ethical rules.
\end{abstract}

 
\section{Introduction} 
\label{sec:introduction}
\input{icseseis-introduction}

\section{Background} 
\label{sec:background}
\input{icseseis-background}

\section{\rev{Codes of Conduct in OSS projects}}
\label{sec:flags}
\input{icseseis-flags}

\section{Our Approach}
\label{sec:approach}
\input{icseseis-approach}

\section{Tool Support}
\label{sec:tool}
\input{icseseis-tool}

\section{Related Work}
\label{sec:related}
\input{icseseis-related}

\section{Conclusions}
\label{sec:conclusion}
\input{icseseis-conclusion}

\section*{Acknowledgments}
This work is part of the project TED2021-130331B-I00 funded by MCIN/AEI/10.13039/501100011033 and European Union NextGenerationEU/PRTR, and the research network RED2022-134647-T (MCIN/AEI/10.13039/501100011033).

\bibliographystyle{IEEEtran}
\bibliography{ms}

\end{document}

%% file: icseseis-introduction.tex
Open-Source Software (OSS) leverages the collaborative work of contributors to keep software projects alive and evolving.
As with any other collaborative endeavor, the development of OSS projects not only relies on technical skills but also on effective communication and social interactions among the contributors of the project.

The potentially rich and diverse mixture of contributors in OSS projects makes this social aspect challenging, as it involves addressing issues related to ethical aspects such as respect, inclusion, and diversity.
To this end, OSS projects have been increasingly deploying a declaration of their code of conduct, which defines ethical rules to ensure a respectful and inclusive participatory environment in the community.
Implementing and fostering codes of conduct in OSS projects not only improves community management and behavior~\cite{DBLP:conf/wcre/TouraniAS17, DBLP:conf/icse/VasilescuFS15, Hsieh2023}, but also establishes a more inclusive and respectful environment for all contributors~\cite{DBLP:journals/pacmhci/LiPFD21, Ehmke2018}.

In the last few years, several initiatives have focused on facilitating the definition of codes of conduct in OSS projects, being the Contributor Covenant~\cite{ContributorCovenant} one of the most recognized proposals.
The Contributor Covenant is a code of conduct that covers common and agreed ethical rules to ensure a respectful and inclusive environment in OSS projects, and usually recommended for OSS projects~\cite{DBLP:journals/sqj/SinghBB22}.
Currently managed by the Organization for Ethical Source\footnote{\url{https://ethicalsource.dev/}}, \rev{the Contributor Covenant has been adopted by a number of relevant OSS projects, such as Angular, Kubernetes, and React}. 
However, we argue that the broad adoption of codes of conduct in OSS projects is still limited, as we will analyze in this paper; and its enforcement is even more challenging, as simply adopting a code of conduct will not prevent conflict or toxicity in OSS communities.

The development of software projects generally involves the use of automation tools or bots to support the management of the project, ranging from continuous integration to code review tools~\cite{Shihab2022}.
In the context of OSS projects, the use of these tools mainly aims to cover the lack of resources and time of the contributors.
Thus, there is a growing trend in the use of bots to automate tasks such as code reviews~\cite{Lin2024}, pull requests~\cite{Wessel2020} or dependency management~\cite{ErlenhovNL22}.

In this paper, we propose a bot-based approach to effectively manage codes of conduct in OSS projects based on the Contributor Covenant.
To motivate our approach, we first present a study on the presence of codes of conduct in OSS projects, and their alignment with the Contributor Covenant.
We then describe our approach to managing codes of conduct, which includes support for defining, monitoring, and enforcing codes of conduct in OSS projects.
Our solution relies on several bots that assist OSS communities in managing the code of conduct of their projects.

The rest of the paper is structured as follows.
Section~\ref{sec:background} describes the background and motivation. 
Section~\ref{sec:flags} studies the presence of code of conduct and presents the main ethical flags to address.
Section~\ref{sec:approach} describes our approach, and Section~\ref{sec:tool} the developed tool support.
Finally, Sections~\ref{sec:related} and~\ref{sec:conclusion} present the related work and the conclusions, respectively.

%% file: icseseis-background.tex
\rev{In this section, we first introduce the use of bots in OSS projects and then we provide an overview of the main concepts related to ethical issues in software development}.

\subsection{Bots in Open-Source Software Development}
Bots have become key in OSS projects to automate tasks like code reviews, issue triaging, and governance, reducing contributors' workloads. % and allowing them to focus on complex tasks. 
Bots respond to events triggered by tools or messages, working as interfaces between users and services, supporting both technical and social activities, including communication and decision-making~\cite{Shihab2022}. 
In OSS, bots are notably useful in tackling sustainability challenges by automating repetitive tasks and improving project efficiency~\cite{Lin2024}.\looseness-1

Several bots are widely adopted in OSS projects. 
Some well-known bots are 
(1) Mergify\footnote{\url{https://mergify.com/}}, which automates the integration of pull requests when they meet certain conditions, reducing the manual effort required; 
(2) Probot\footnote{\url{https://probot.github.io/}}, which allows for the creation of customized bots for project-specific tasks, such as enforcing code standards;
or (3) Dependabot\footnote{\url{https://github.com/dependabot}}, which helps keep projects up to date by managing dependencies. %, while Stale Bot assists by closing inactive issues, ensuring the project remains focused on current tasks.

\subsection{Ethical Issues in Open-Source Software Development}
\label{sec:background:ethical}
The development of OSS software inherently faces several ethical challenges, which can be categorized into various types. 
One of the primary concerns is inclusion and diversity, where projects aim to involve contributors from varied cultural and demographic backgrounds~\cite{Damian2024,Albusays2021}. 
Ensuring equitable access and participation is crucial for fostering a diverse community, but it also opens up ethical dilemmas related to bias, exclusionary practices, and discrimination~\cite{McIlwain2019}.

Another prominent issue is fairness and credit attribution. 
OSS thrives on contributions from volunteers, but often contributions may not be adequately recognized, leading to dissatisfaction or unfair treatment of contributors. 
Ethical dilemmas arise when project leaders fail to properly acknowledge or attribute work, causing resentment and mistrust within the community~\cite{Young2021,Casari2021,HippocraticLicense2021}.

Power dynamics also play a critical role in ethical discussions, as certain contributors or maintainers may hold disproportionate influence over project decisions. 
This can lead to scenarios where the decision-making processes are opaque or exclusive, marginalizing less influential contributors and limiting community growth~\cite{Finley2022,Maenpaa2018,Buritica2019,FariasICSOB2021}.

Additionally, conflict resolution and governance represent significant ethical concerns. 
OSS communities often operate without formalized hierarchies or governing bodies, meaning conflicts can escalate quickly, especially when clear protocols are not in place. 
This is where codes of conduct become vital, as they offer guidance for acceptable behavior, ensure accountability, and help manage these disputes fairly and transparently~\cite{DBLP:conf/wcre/TouraniAS17, DBLP:journals/sqj/SinghBB22, DBLP:journals/pacmhci/LiPFD21, Ehmke2018}. 
These ethical concerns, specifically related to the social dynamics of diverse OSS communities, make it essential to establish guidelines, such as a code of conduct, to mitigate risks and foster an inclusive environment.

%% file: icseseis-flags.tex
Despite the importance of codes of conduct, we argue that their presence in OSS projects is still very limited. 
To assess the relevance and structure of existing codes of conduct, we perform a study to analyze their existence in OSS projects and their resemblance to the Contributor Covenant. 
We then define of a set of ethical flags based on the Contributor Covenant and study their presence in current codes of conducts\footnote{The replicability package of our studies can be found at: \url{http://hdl.handle.net/20.500.12004/1/C/ICSESEIS/2025/883}.}. 

\subsection{\rev{Presence of Codes of Conduct in OSS}} 
\label{sec:background:cocoss}
\rev{To assess current practices in OSS projects, we have conducted a study on the presence of codes of conduct on GitHub. To this end, we analyzed the presence of code of conduct files in OSS projects hosted on GitHub and labeled them according to the main programming languages they use.}

We studied the 1,000 top-starred repositories for 12 different programming languages, namely: C, C++, C\#, Go, Java, JavaScript, PHP, Python, Ruby, Rust, Scala, and TypeScript. 
We analyzed a total of 12,000 repositories.
To perform the study, we first retrieved the list of the top 1,000 starred repositories for each programming language using the GitHub API.
Then, for each repository, we check the presence of the file corresponding to the code of conduct via the GitHub API.
The search is done by looking for the presence of a file named \texttt{CODE\_OF\_CONDUCT.md} (and all possible variations, such as \texttt{coc.md}, \texttt{code-of-conduct.txt}, \texttt{CodeOfConduct.adoc}), in the typical locations of the repository. 
The second and third columns of Table~\ref{tab:cocAnalysis} show the results of this search. 
As can be seen, the presence of a code of conduct is scarce, with an average of 19.32\% of the repositories having a code of conduct file.
We also checked that the length of the file to spot files too short to be considered a code of conduct.
\rev{ Table~\ref{tab:cocAnalysis} shows the number and percentage of projects with a code of conduct file containing more than 5 lines of text. This threshold is used to ensure the file is not a placeholder or a link with minimal context, as short files often redirect to external resources that may contain the actual code of conduct. Further analysis revealed that only 13.29\% of the analyzed files included a link to an external webpage.}

\begin{table}[t]
  \centering
  \caption{Presence of code of conduct files in OSS projects. CC = Contributor Covenant.}
  \label{tab:cocAnalysis}
  \begin{tabularx}{\columnwidth}{Xrr@{\hspace{0.5em}}r@{\hspace{0.5em}}rrr}
      \multicolumn{1}{c}{\textsc{Programming}} &
      \multicolumn{2}{c}{\textsc{Presence}} & &
      \multicolumn{2}{c}{\textsc{Content}} &
      \multicolumn{1}{c}{$r_{\text{cc}}$} \\ 
      \cmidrule{2-3} \cmidrule{5-6} 
      \multicolumn{1}{c}{\textsc{Language}} &
      \multicolumn{1}{c}{\textsc{\#}} &
      \multicolumn{1}{c}{\textsc{\%}} & & 
      \multicolumn{1}{c}{\textsc{\#}} &
      \multicolumn{1}{c}{\textsc{\%}} &
      \multicolumn{1}{c}{\textsc{\%}} \\
    \toprule
      C          &   101 & 10.10 & &    87 &  8.70 & 80.22 \\ 
      C\#        &   213 & 21.30 & &   178 & 17.80 & 86.93 \\ 
      C++        &   171 & 17.10 & &   148 & 14.80 & 74.19 \\ 
      Go         &   290 & 29.00 & &   208 & 20.80 & 86.33 \\ 
      Java       &   107 & 10.70 & &    94 &  9.40 & 80.20 \\ 
      JavaScript &   242 & 24.20 & &   215 & 21.50 & 87.78 \\ 
      PHP        &   117 & 11.70 & &   110 & 11.00 & 86.61 \\ 
      Python     &   223 & 22.30 & &   193 & 19.30 & 78.32 \\ 
      Ruby       &   190 & 19.00 & &   176 & 17.60 & 86.59 \\ 
      Rust       &   230 & 23.00 & &   202 & 20.20 & 83.73 \\ 
      Scala      &   109 & 10.90 & &    97 &  9.70 & 47.52 \\ 
      TypeScript &   325 & 32.50 & &   302 & 30.20 & 90.58 \\ 
    \midrule
      Total      & 2,318 & 19.32 & & 2,010 & 20.10 & 82.85 \\  
    \bottomrule 
  \end{tabularx}
\end{table}

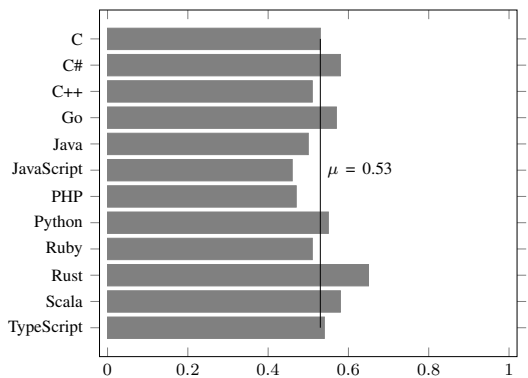
\begin{figure}[t]
  \centering
\resizebox{0.8\columnwidth}{!}{%
\begin{tikzpicture}
  \begin{axis}[
    xbar, 
    enlarge y limits  = 0.1,
    enlarge x limits  = 0.02,
    ytick             = data,
    symbolic y coords = {TypeScript, Scala, Rust, Ruby, Python, PHP, JavaScript, Java, Go, C++, C\#, C},
    label style       = {font=\footnotesize},
    tick label style  = {font=\footnotesize},
    legend style      = {font=\footnotesize},
    xmin              = 0,
    xmax              = 1,
    legend style      = {at={(0.5, -0.08)},anchor=north, legend columns=-1},
    minor x tick num = 0,
  ]
  \addplot [style={gray,fill=gray,mark=none}] coordinates { 
    (0.53,C)    
    (0.58,C\#)    
    (0.51,C++)    
    (0.57,Go)    
    (0.50,Java)   
    (0.46,JavaScript)  
    (0.47,PHP)  
    (0.55,Python)   
    (0.51,Ruby)   
    (0.65,Rust)   
    (0.58,Scala)   
    (0.54,TypeScript) 
  };
  \addplot [black,line legend, sharp plot,update limits=false] coordinates {
    (0.53,C)    
    (0.53,C\#)      
    (0.53,C++)        
    (0.53,Go)     
    (0.53,Java)     
    (0.53,JavaScript)     
    (0.53,PHP)      
    (0.53,Python)      
    (0.53,Ruby)       
    (0.53,Rust)       
    (0.53,Rust)       
    (0.53,TypeScript)    
  }; 
  \node[text width=2cm] at (70,60) {\footnotesize $\mu$ = 0.53};
  \end{axis}
\end{tikzpicture}
}
    \caption{Freshness of Code of Conduct ($f_{\text{CoC}}$) across different programming languages, \rev{with values from 0, which indicates the most recent update; to 1, which indicates the repository's creation date.}}
    \label{fig:f_coc}
\end{figure}

We also studied the last time codes of conduct were modified.
To this aim, we introduced the concept of freshness of a code of conduct or $f_{\text{CoC}}$, defined as $f_{\text{CoC}} = \frac{t_{\text{CoC}}}{l_{\text{repo}}}$ where $t_{\text{CoC}}$ is the timespan in years between the last modification of the code of conduct and the last commit, and $l_{\text{repo}}$ is the lifespan in years of the repository.
Thus, an $f_{\text{CoC}}$ of value 1 implies that the code of conduct was last modified at the creation of the repository, whereas a ratio of 0 indicates it was last updated at the time of the most recent commit.
The average ratio across all repositories is 0.53 (see the sixth column of Figure~\ref{fig:f_coc}), suggesting that modifications to the code of conduct tend to occur roughly halfway through the repository's lifespan. 
This result reveals that updates to the code of conduct may not be needed as frequently as other parts of the repository.

Finally, we studied the alignment of the collected codes of conduct with the Contributor Covenant. 
We analyzed the references from the collected codes of conduct to the Contributor Covenant, which we refer to as $r_{\text{cc}}$.
The last column of Table~\ref{tab:cocAnalysis} shows the results of this analysis.
The results reveal that, on average, 82.85\% of the collected codes of conduct refer to the Contributor Covenant.
Further analysis conducted by searching for verbatim phrases from specific versions of the Contributor Covenant revealed that 55.18\% of them included sentences from Contributor Covenant v1.4, while 26.78\% of them contained phrases of versions 2.0 or 2.1. % of the Contributor Covenant.

All in all, our analysis confirms that the adoption of codes of conduct in OSS projects is low, with their presence in only 19.32\% of the analyzed repositories.
Furthermore, when present, the analyzed codes of conduct are mainly updated halfway through the repository's lifespan, and most of them refer to the Contributor Covenant, even including verbatim sentences of specific versions of the Contributor Covenant.

\subsection{\rev{Ethical Flags from the Contributor Covenant}}
\rev{Given the evidence found on the alignment of the collected codes of conduct with the Contributor Covenant (both as references and the use of verbatim phrases), we have defined a set of behavioral markers, or flags, to define a reference set of ethical behaviors in OSS. We will use these flags later in our approach to managing codes of conduct in OSS projects. }

\rev{To identify the ethical flags, we focused on their relevance to common behavioral patterns in OSS communities and their importance in fostering a positive and inclusive environment. The ten identified ethical flags were derived directly from the ``Standards'' section of the version 2.1 of Contributor Covenant, which provides actionable guidelines for community behavior.}
We categorized the flags into two groups: positive flags, which represent desirable behaviors; and negative flags, which denote harmful or inappropriate behaviors in community settings. 
Table~\ref{tab:flags} lists the flags identified. 
The table includes the flag ID, name, description, and examples of keywords identifying the flag in the text of codes of conduct.

\begin{table*}[t]
    \renewcommand{\cellalign}{cc}
    \renewcommand{\theadalign}{cl}
    \renewcommand{\arraystretch}{2.0}
    \centering
    \caption{Ethical Flags from the Contributor Covenant}
    \label{tab:flags}
    \begin{tabularx}{\textwidth}{c@{\hspace{0.75em}}c@{\hspace{0.5em}}l@{\hspace{0.5em}}l@{\hspace{0.5em}}l}
        \toprule
            &
            \multicolumn{1}{c}{\textsc{Id}} & 
            \multicolumn{1}{c}{\textsc{Name}} & 
            \multicolumn{1}{c}{\textsc{Description}} &
            \multicolumn{1}{c}{\textsc{Keyword Examples}} \\
        \midrule
        \multirow{5}{*}{\rotatebox[origin=c]{90}{Positive\hspace{1em}}} & 
          F1  & Empathy and Kindness             & Demonstrating understanding and compassion towards others              & \makecell{``being kind'', ``empathy'', \\ ``empathic'', ``kindness''} \\
        & F2  & Respect for Differences          & Valuing diverse perspectives and backgrounds                           & \makecell{``how respect'', ``be respectful'', \\ ``eliminate biases''}\\
        & F3  & Constructive Feedback            & Providing feedback that is helpful and aimed at improvement            & \makecell{``respectful criticism'', \\ ``thoughtful addressing''}\\
        & F4  & Responsibility and Apology       & Taking responsibility for one's actions and apologizing when necessary & \makecell{``apologize'', ``admit fault'', \\ ``responsibility and apologizing''} \\
        & F5  & Common Good                      & Acting in ways that benefit the broader community                      & \makecell{``best for the community'', \\ ``community benefit''}\\
        \midrule
        \multirow{5}{*}{\rotatebox[origin=c]{90}{Negative\hspace{1em}}} & 
          F6  & Sexualized Language or Imagery   & Using language or imagery that is inappropriate and sexual in nature   & \makecell{``unwelcome sexual attention'', \\ ``sexual language'', ``sexual jokes''}\\
        & F7  & Insulting or Derogatory Comments & Making comments that insult or demean others                           & \makecell{``derogatory comments'', ``trolling'', \\ ``ridicule'', ``insults''}\\
        & F8  & Public or Private Harassment     & Engaging in behavior that intimidates or harasses others               & \makecell{``threatened'', ``harassing'',\\ ``stalking'', ``bullying''}\\
        & F9  & Publishing Private Information   & Sharing private information about others without consent               & \makecell{``doxing'', ``privacy breach'', \\ ``unconsented disclosure''}\\
        & F10 & Inappropriate Conduct            & Behaving in a manner that is not suitable in a professional setting              & \makecell{``unsuitable behavior'', \\ ``behaving professionally''}\\
        \bottomrule
    \end{tabularx}
\end{table*}

These flags provide a comprehensive framework for evaluating the ethical dimensions of OSS projects.
They enable the analysis of existing codes of conduct (and their alignment with the Contributor Covenant), thus identifying potential dimensions for improvement.
Moreover, the flags can be used to monitor the behavior of contributors in OSS projects, providing a mechanism for detecting and addressing harmful or inappropriate behaviors.

\subsection{\rev{Flag Usage in Existing Codes of Conduct}}
We studied the presence of the identified ethical flags in the codes of conduct of the collected OSS projects. 
To perform this study, we applied Natural Language Processing techniques to clean, lemmatize, and analyze the content of each code of conduct, comparing it against a predefined dictionary of keywords (as shown in the last column of Table~\ref{tab:flags}) to identify the ethical flags.
\rev{The most commonly used codes of conduct in OSS projects were analyzed to generate the dictionary of keywords, incorporating synonyms and related expressions to enhance its coverage. This dictionary includes non-overlapping keywords for each ethical flag.}
\rev{This approach follows a methodology similar to that often used in sentiment analysis, where lexicon-based techniques provide a controlled and structured method for categorizing specific indicators within a text~\cite{nandwani2021,imran2022}. In this case, the methodology is adapted to detect keywords in codes of conduct, using a predefined lexicon to maintain fous on the unique characteristics of each ethical flag, much like sentiment lexicons are designed to identify specific emotions in text analysis.}
Table~\ref{tab:flags_by_language} shows the results.

\begin{table*}[t]
    \centering
    \caption{Percentage of Flags in Code of Conduct Files by Language}
        \begin{tabularx}{0.7\textwidth}{Xccccccccccc}
            \toprule
              \multicolumn{1}{c}{\textsc{Language}} & 
              \multicolumn{1}{c}{\textsc{F1}} & 
              \multicolumn{1}{c}{\textsc{F2}} & 
              \multicolumn{1}{c}{\textsc{F3}} & 
              \multicolumn{1}{c}{\textsc{F4}} & 
              \multicolumn{1}{c}{\textsc{F5}} & 
              \multicolumn{1}{c}{\textsc{F6}} & 
              \multicolumn{1}{c}{\textsc{F7}} & 
              \multicolumn{1}{c}{\textsc{F8}} & 
              \multicolumn{1}{c}{\textsc{F9}} & 
              \multicolumn{1}{c}{\textsc{F10}} \\
            \midrule
            C          & 89.16 & 96.39 & 87.95 & 32.53 & 89.16 & 95.18 &  98.80 &  96.39 & 93.98 & 87.95 \\
            C\#        & 91.41 & 93.75 & 91.41 & 23.44 & 91.41 & 98.44 & 100.00 & 100.00 & 99.22 & 91.41 \\
            C++        & 86.61 & 93.70 & 83.46 & 28.35 & 81.89 & 93.70 &  96.85 &  99.21 & 91.34 & 83.46 \\
            Go         & 88.78 & 89.27 & 84.88 & 32.20 & 84.88 & 93.66 &  94.63 &  98.05 & 92.20 & 84.88 \\
            Java       & 85.23 & 89.77 & 85.23 & 20.45 & 85.23 & 96.59 &  97.73 &  98.86 & 95.45 & 84.09 \\
            JavaScript & 90.24 & 90.73 & 86.34 & 22.44 & 86.83 & 96.59 &  98.05 &  99.51 & 92.68 & 86.83 \\
            PHP        & 81.31 & 89.72 & 83.18 & 19.63 & 83.18 & 89.72 &  96.26 &  94.39 & 89.72 & 81.31 \\
            Python     & 93.10 & 93.10 & 87.93 & 36.21 & 85.63 & 93.10 &  94.83 &  98.28 & 91.38 & 88.51 \\
            Ruby       & 69.64 & 72.02 & 69.64 & 15.48 & 68.45 & 95.24 &  97.02 &  99.40 & 85.12 & 68.45 \\
            Rust       & 96.30 & 84.66 & 84.13 & 51.32 & 83.60 & 84.13 &  96.83 &  98.94 & 84.66 & 83.07 \\
            Scala      & 64.56 & 62.03 & 58.23 & 13.92 & 58.23 & 63.29 &  64.56 & 100.00 & 59.49 & 58.23 \\
            TypeScript & 91.44 & 92.47 & 90.41 & 35.96 & 88.36 & 95.89 &  98.29 &  99.66 & 94.18 & 89.38 \\
            \midrule
            \textbf{Total} & \textbf{85.65} & \textbf{87.30} & \textbf{82.73} & \textbf{27.66} & \textbf{82.24} & \textbf{91.29} & \textbf{94.49} & \textbf{98.56} & \textbf{89.12} & \textbf{82.30} \\
            \bottomrule
        \end{tabularx}
    \label{tab:flags_by_language}
\end{table*}

\rev{The results reveal several key insights. Firstly, there is a broad adoption of all positive flags except for F4, with an adoption significantly lower (27.66\%). This discrepancy is particularly notable because F4, which is related to positive reinforcement behaviors, was introduced in version 2.0 of the Contributor Covenant. This suggests that many projects have not updated their codes of conduct to align with the latest standards, potentially highlighting an improvement point to foster constructive interactions within communities. Moreover, the high presence of flags related to negative behaviors, such as F7 (94.49\%) and F8 (98.56\%), may indicate that most projects prioritize preventing harmful behaviors.}

We believe that these findings also highlight the importance of reviewing frequently and maintaining up-to-date codes of conduct.
Our approach aims to address this challenge by automating the process of checking for outdated or incomplete codes of conduct and suggesting improvements based on the latest versions of ethical guidelines. 

\subsection{\rev{Threats to Validity}}
\rev{We acknowledge our studies are subjected to a number of threats to validity. Regarding the internal validity (i.e., inferences we have made), we relied on the data provided by the GitHub API, which may not be complete and/or include toy projects (e.g., projects addressing homework assignments). To minimize this threat, we analyzed a large set of top repositories of each programming language. Regarding the external validity, note that our data was collected on June 19$^{\text{th}}$, 2024; and therefore our results should not be directly generalized without proper comparison and validation.}

%% file: icseseis-approach.tex
We propose an approach to effectively managing codes of conduct in OSS projects.
Our approach relies on a bot-based solution that helps with the tasks of reacting, monitoring, analyzing, and initializing codes of conduct. 
Each task is addressed by a specific bot, and thus we defined four bots; which are orchestrated by an additional bot.

\begin{figure*}
  \includegraphics[width=\textwidth]{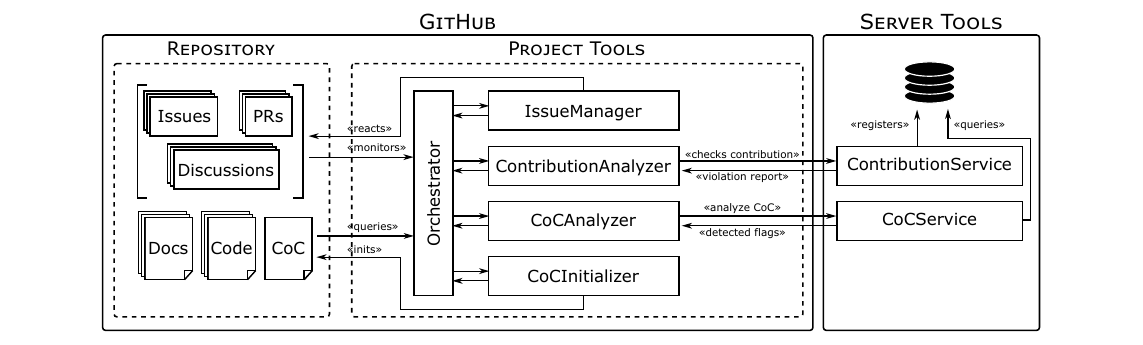}
  \vspace{-0.5cm}
  \caption{Architecture of our approach.}
  \label{fig:approach}
\end{figure*}

Figure~\ref{fig:approach} illustrates the main elements of our approach, which include three groups:
(i) the GitHub repository, containing code and documentation; and the corresponding collaboration elements, that is, a set of issues, pull requests (PR) and discussions (and the comments they may have);
(ii) the project-related supporting tools, which include the devised bots and provides specific support for monitoring the collaboration elements, analyzing the code of conduct, and initializing it;
and (iii) the server-side supporting tools, providing support to the bots requiring specific computational resources.

In the following, we describe how we address the tasks covered by our approach, grouped into three main functionalities: 
(1) analyzing codes of conduct, which include bots for analyzing and initializing codes of conduct;
(2) monitoring in OSS projects, which cover bots to monitor and react to contributions;
and (3) orchestration, which coordinates the bots and ensures that the tasks are orderly performed.

\subsection{Analyzing Codes of Conduct}
Our approach first analyzes the code of conduct in the OSS project by checking whether the corresponding file exists.
If the file is missing, the orchestrator triggers the bot \emph{CoCInitializer} to create one \rev{by submitting a pull request to the repository with version 2.1 of the Contributor Covenant}; otherwise, the bot \emph{CoCAnalyzer} is activated to review the file's structure and content. 
Once the file has been validated, the code of conduct is submitted to the \emph{CoCService} for detailed analysis.

The \emph{CoCService} then examines the content of the code of conduct to identify which ethical flags are present, thus providing the \emph{CoCAnalyzer} bot with a summary that includes the detected flags and the specific version of the code of conduct in use.

If important flags are missing or the code of conduct file is outdated, the orchestrator launches an additional automated process to resolve the issue.
In these cases, the bot \emph{IssueManager} generates issues that notify maintainers about the missing elements, detailing which positive or unacceptable behavior flags are absent. 
This automated process allows maintainers to review and integrate the proposed improvements, ensuring the project follows an up-to-date ethical framework.

\subsection{Monitoring in OSS projects}
Once the code of conduct of the project has been analyzed and the project's flags have been identified, the orchestrator monitors contributions in OSS projects, and launches the bot \emph{ContributionAnalyzer} to perform their analysis when they are created.
The bot \emph{ContributionAnalyzer} relies on the service \emph{ContributionService}, which leverages a natural language model to analyze the contributions and assess whether they comply with the community's established code of conduct (i.e., detected ethical flags). 
Based on the analysis, the bot determines if the contribution is positive, negative, or neutral.

For positive contributions, the bot reacts and posts an automatic response in the corresponding repository event. 
If the analysis detects a code of conduct violation, \rev{the bot proceeds to mark the comment.} 
If the contribution is neutral, nothing is done.
Additionally, the bot sends notifications to project administrators via email, providing details about the actions taken, and the content analyzed. 
All information related to the contribution (i.e., analysis and actions executed) is stored in a database, enabling auditing and event-tracking. 

\subsection{Orchestration} 
\begin{figure}[t]
    \includegraphics[width=\columnwidth]{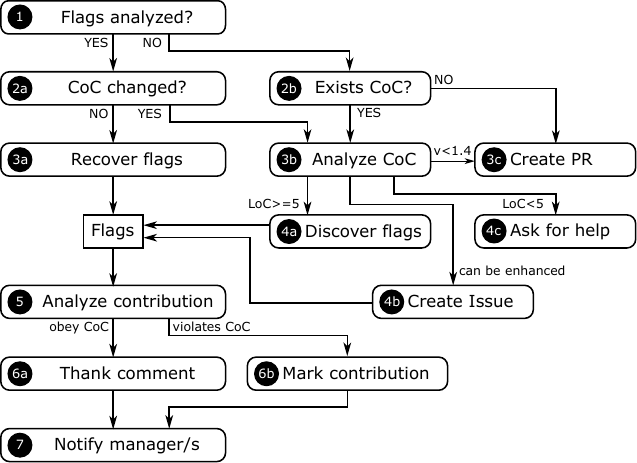}
    \caption{\rev{Process managed by the Orchestrator bot.}}
    \label{fig:orchestration} 
        
\end{figure}

The Orchestrator is the bot responsible for managing the interactions between the various modules of the tool, triggering other bots and services when certain conditions are met. 
The orchestration process is illustrated in Figure~\ref{fig:orchestration}, which we describe in the following.

Each time a contribution (i.e, discussion, issue, comment, or pull request) is created, the orchestrator first checks if the ethical flags of the project have already been analyzed (cf. \circled{1}).
If the flags have been analyzed, the process checks if the code of conduct has changed since then (cf. \circled{2a}).
On the other hand, if the flags were not analyzed before, the process checks if the code of conduct exists (cf. \circled{2b}).
If the code of conduct was analyzed and has not changed, the previously identified flags of the contribution are recovered (cf. \circled{3a}).
Whether the code of conduct has changed or exists, it is analyzed to (re-)discover the flags (cf. \circled{3b}).
Finally, if the code of conduct does not exist, a first version is proposed via pull request in the project (cf. \circled{3c}).

The analysis of the code of conduct can drive to four possible actions: 
(1) if the code of conduct is present and meets the minimum length requirement \rev{(currently configured to LoC $\geq$ 5), ensuring it is not merely a placeholder or a link with minimal context,} the tool proceeds to discover the ethical flags within the code of conduct (cf. \circled{4a}),
(2) if the code of conduct does not cover all ethical flags, the flags are discovered but an issue to discuss an enhancement of the rules is also proposed (cf. \circled{4b}),
(3) if the code of conduct is present but does not meet the minimum length requirement (LoC $\geq$ 5), the system prompts for help  (cf. \circled{4c}),
and (4) if a code of conduct is found, but it is based on an outdated version of the Contributor Covenant (i.e., version older than v1.4), the tool creates a pull request to propose \rev{version 2.1 of the Contributor Covenant.} {(cf. \circled{3c}). 

Once the flags have been retrieved (from a previously analyzed contribution via \circled{3a} or a new one via \circled{4a}), the \emph{Orchestrator} triggers the analysis of the contribution (cf. \circled{5}), comparing the behavior exhibited in the contribution with the ethical guidelines defined in the code of conduct.
Based on the results of the contribution analysis, one of two actions is taken:
(1) if the contribution complies with the code of conduct, the system generates an automatic thank-you comment (cf. \circled{6a}); otherwise, \rev{the system proceeds to mark the contribution as inappropriate. (cf. \circled{6b}) }
Finally, the \emph{Orchestrator} notifies the project maintainers about the actions taken (cf. \circled{7}), providing a detailed report of the analysis and any violations detected.

This orchestrated process ensures that every contribution and every code of conduct is monitored and managed automatically, improving the efficiency of moderation and the enforcement of ethical standards in OSS projects.

%% file: icseseis-tool.tex
In this section, we describe the implementation of our tool, which integrates with GitHub workflows to enforce and manage codes of conduct\footnote{The implementation can be found at: \url{http://hdl.handle.net/20.500.12004/1/C/ICSESEIS/2025/032}.}.
The overall architecture, depicted in Figure~\ref{fig:approach}, leverages GitHub Actions, external analysis services, and automation scripts to continuously observe the state of the repository and react accordingly.
Next we describe the implementation details of the tool, including the key components and the use cases that illustrate the tool's functionality.

\subsection{Implementation}
\label{sec:implementation}
The implementation of the tool is divided into several key components, namely: 
(1) GitHub Action bots, (2) server-side services, and (3) database and persistence.
Next we describe the main implementation details of each component. 

\subsubsection{GitHub Actions Bots}
Bots have been implemented using GitHub Actions, allowing seamless integration with repository events (e.g., issues or pull creation events). 
Each bot is designed as an independent action within GitHub workflows. 
Thus, we implemented five bots, which aligned with those shown in Figure~\ref{fig:approach}: \emph{CoCAnalyzer}, \emph{ContributionAnalyzer}, \emph{CoCInitializer}, \emph{IssueManager}, and \emph{Orchestrator}. 
This modular design ensures a distribution of responsibilities and promotes maintainability and scalability.

To implement the bots, we relied on the support for scripting via Bash commands. 
The use of Bash scripts within the workflows allows us to implement advanced conditional logic directly in continuous integration processes. 
Moreover, it allows us to make the workflows lighter, thus eliminating the need for additional or external tools.
For example, using Bash scripts, we have automatized the analysis of behavior patterns in comments and contributions within the GitHub Action execution in real time. 
Additionally, the reuse of logic across bots is straightforward, which simplifies extending functionalities without the need to rewrite large portions of the code.
Finally, Bash scripts facilitate the handling of complex data and integration with external APIs. 
For example, in the \texttt{CoCAnalyzer} bot, Bash scripts are used to process the \texttt{CODE\_OF\_CONDUCT.md} file, validate it, and send it for external analysis.
Listing~\ref{lst:example} shows an example that demonstrates how \texttt{jq} converts the \texttt{CODE\_OF\_CONDUCT.md} file into a JSON object for structured analysis, \texttt{curl} sends it to a remote server for processing, and \texttt{jq} processes the server's response, extracting flags or indicators to trigger automatic actions.

\begin{lstlisting}[language=Bash, float=t, style=codeStyle, label=lst:example, caption=Example of Bash script used to implement the bots.]
# Convert the content of CODE_OF_CONDUCT.md to JSON
content=$(cat ${{ env.coc_file }})
repo_name="${{ env.repo_name }}"
repo_url="${{ env.repo_url }}"
jq -n --arg type "code_of_conduct" 
      --arg code_of_conduct "$content" 
      --arg repository_name "$repo_name" 
      --arg repository_url "$repo_url" \
'{
  type: $type,
  code_of_conduct: $code_of_conduct,
  repository_name: $repository_name,
  repository_url: $repository_url
}' > data/payload.json

# Send to server for analysis
response=$(curl -s -X POST "${{ inputs.SERVER }}" 
                -H "Content-Type: application/json" 
                --data @data/payload.json)
echo "$response" > data/response.json

# Check the response status and extract flags
status=$(jq -r '.data.status' data/response.json)
flags=$(jq -c '.data.flags' data/response.json)
echo "Flags: $flags"
\end{lstlisting}

\subsubsection{Server-Side Services}
The server-side services includes two main components, as shown in Figure~\ref{fig:approach}: the \emph{CoCService} and the \emph{ContributionService}. These services are responsible for performing intensive processing tasks that cannot be executed directly in GitHub Actions.

The \emph{CoCService} applies the process described in Section~\ref{sec:flags} to identify ethical flags in the code of conduct, and determines the version of the Contributor Covenant in use, by looking for exact phrases of each version.
The \emph{CoCService} then returns a JSON object containing the detected flags and version details, enabling the \emph{CoCAnalyzer} bot to take actions that ensure the code of conduct is monitored and enforced effectively.

The \emph{ContributionService} performs two main functions using LLMs:
(1) sentiment analysis on contributions to determine if they align with the community's established code of conduct (i.e., ethical rules detected in the code of conduct);
and (2) generation of automatic responses when a contribution adheres to the defined positive flags.

To carry out effective sentiment analysis focused on the identified ethical flags, we rely on the use of LLMs.
These models have proven to outperform traditional systems in emotional classification, which is critical for detecting complex nuances such as irony and sarcasm~\cite{Carneros2023}. 
In our context, this is particularly relevant, as the ethical flags contain emotional subtleties that are challenging to assess with conventional sentiment analysis models. 
Additionally, LLMs eliminate the need for extensive preprocessing, enabling a more streamlined and direct analysis of comments~\cite{Nadi2024}.
In our approach, we deployed the Mixtral-8x22B model locally using Ollama\footnote{\url{https://ollama.com/emsi/mixtral-8x22b}}.

We developed a prompt specifically designed to align with the predefined flags that facilitates effective sentiment analysis. 
Its structure instructs the system to analyze community comments and generates a JSON output that includes the comment, its classification (positive, negative, or neutral), reasons for the classification, and a list of relevant flags based on the ones we have identified. 
The input prompt incorporates the code of conduct guidelines, allowing the system to perform a more contextualized and specific analysis of each comment, as shown in Listing \ref{lst:prompt}. 
It also includes five examples (not shown in Listing \ref{lst:prompt} for the sake of space) that reflect the diversity of potential comments and their classifications. 
Listing \ref{lst:json} shows an example of the LLM output.

\begin{lstlisting}[float=t, label=lst:prompt, style=promptStyle, caption={Part of the prompt used for sentiment analysis.}]
The Code of Conduct is based on the following guidelines:

**Positive Flags:**
1. Demonstrating empathy and kindness toward other people
2. Being respectful of differing opinions, viewpoints, and experiences
3. Giving and gracefully accepting constructive feedback
4. Accepting responsibility and apologizing to those affected by our mistakes, and learning from the experience
5. Focusing on what is best not just for us as individuals, but for the overall community

**Negative Flags:**
1. The use of sexualized language or imagery, and sexual attention or advances of any kind
2. Trolling, insulting or derogatory comments, and personal or political attacks
3. Public or private harassment
4. Publishing others private information, such as a physical or email address, without their explicit permission
5. Other conduct which could reasonably be considered inappropriate in a professional setting

**Neutral Flags:**
A comment is considered neutral when it does not fall under any of the described flags.
\end{lstlisting}

\begin{lstlisting}[float=t, label=lst:json, style=codeStyle, numbers=none, caption={Example output from sentiment analysis.}]
{
  "comment": "Thank you for your help, I really appreciate your time and effort.",
  "classification": "positive",
  "reasons": "This comment reflects a positive engagement that fosters a supportive community atmosphere.",
  "flags": ["Demonstrating empathy and kindness toward other people"]
}
\end{lstlisting}

To evaluate the precision of our approach, we used ChatGPT-4o to generate a diverse dataset of text-based comments. 
The comments varied in tone (i.e., direct, subtle, ironic, sarcastic, and some with emoticons) to reflect different communication styles.  
We generated 100 entries for each ethical flag, along with 100 neutral comments as a control group, totaling 1,100 comments.
These entries were manually validated by the authors of this paper, and only 3.10\% of the generated entries were modified to be aligned with the corresponding flag.
Once validated, they were processed through the \emph{ContributionService} for ethical flag detection.
We first validated the performance of the Mixtral model in classifying comments as positive (i.e., F1 to F5), negative (i.e., F6 to F10), or neutral (i.e., no flags), achieving an accuracy of 95.55\%.
\rev{Although the model demonstrates high precision across all classes, we detected that neutral contributions tend to be classified as positive.}
\rev{We also analyzed the ability to detect specific flags, achieving an overall accuracy of 84.02\%. Lower accuracy levels were observed for F6 (i.e., sexualized language or imagery) and F8 (i.e., public or private harassment), with an accuracy of 72.73\% and 67.00\%, respectively.}
\rev{For F2 (i.e., respect for differences), the accuracy was 46.00\%. This misclassification is not considered critical, as F2 is often associated with F1 or F5, which are three positive flags and result in the same supportive action.} 

\begin{figure}[t]
    \centering
    \includegraphics[width=\linewidth]{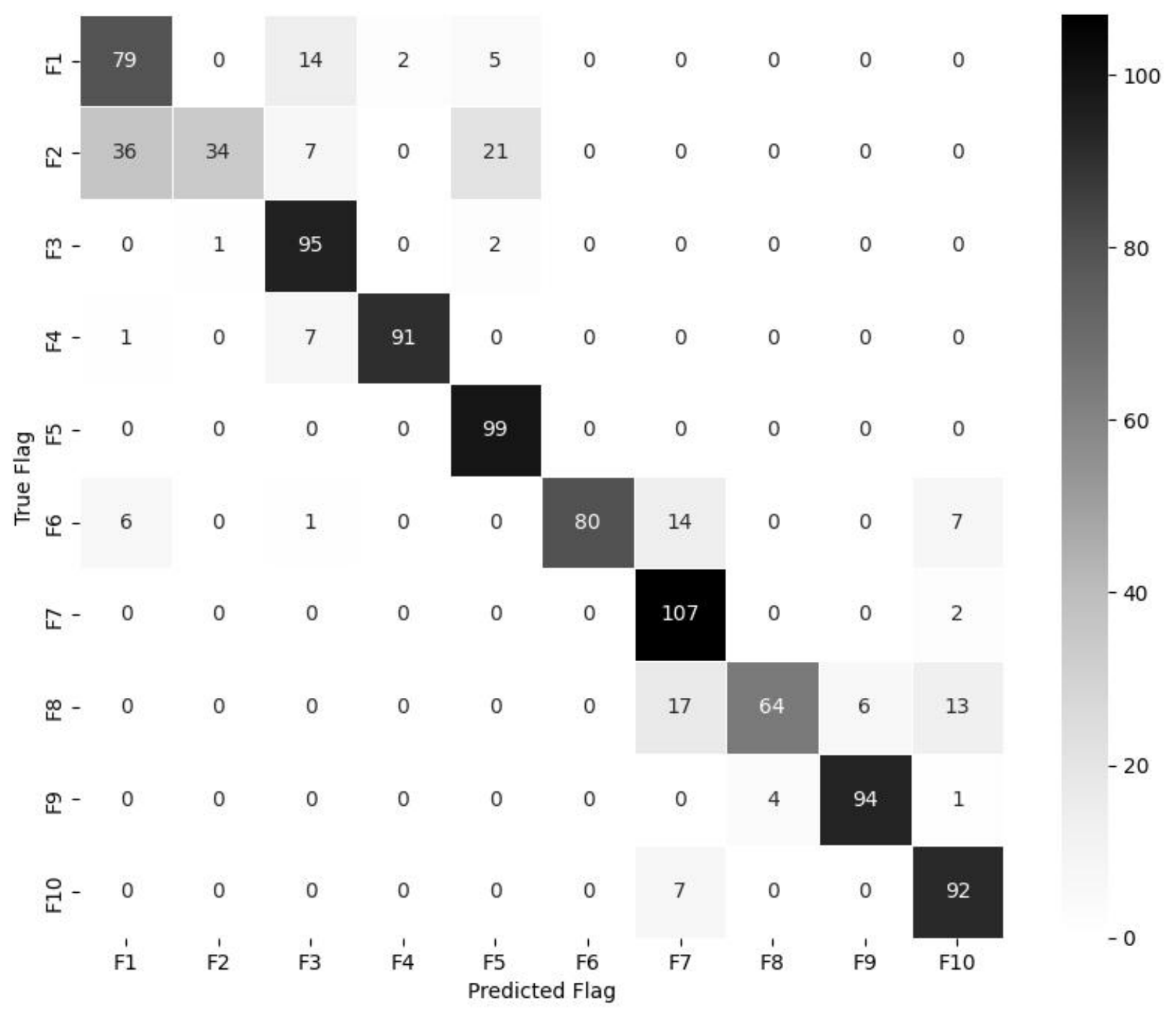}
    \caption{\rev{Confusion matrix illustrating flag detection on the test dataset.}}
    \label{fig:flag_matrix}
\end{figure}

Similarly, F6 overlaps with F7 (i.e., insulting or derogatory comments) and F10 (i.e., inappropriate conduct) due to shared characteristics. 
Additionally, F8, which requires a repetitive pattern, can be interpreted as F7 or F10 because the model may not always detect this pattern in the comments. 
These nuances underscore the complexity of distinguishing between different types of unethical behavior, as illustrated in the confusion matrix shown in Figure~\ref{fig:flag_matrix}, which shows the overlaps and misclassifications among the flags.

When a contribution aligns with positive flags, the \emph{ContributionService} uses an additional specific prompt which instructs the system to create a kind and grateful reply, tailored to the communicative style of the GitHub community. 
The instructions ensure clarity in tone and structure, maintaining a positive focus. 
The resulting reply expresses gratitude and reinforces the community's commitment to collaboration and mutual support, emphasizing the importance of constructive interactions in the OSS environment.

\subsubsection{Database and Persistence}
The results of the contribution and code of conduct analysis are stored in a database to ensure full traceability. 
The database records information such as the repository name, analysis results, detected flags, and timestamps of the analysis. 
This guarantees that maintainers can access a complete history of ethical violations and changes to the code of conduct, enabling a more informed decision-making process for managing community behavior.

Our architecture includes two separate database models, each tailored to the specific needs of the services: one for the \emph{ContributionService} and another for the \emph{CoCService}. 
This division allows for a clear separation of concerns, ensuring that each service can scale independently while maintaining the integrity of its data. 
The \emph{ContributionService} database focuses on storing detailed records of user contributions, including comment analysis, detected behavioral flags, and any automatic responses generated. 
Meanwhile, the \emph{CoCService} database stores information about the project's code of conduct, such as its content, version history, and any detected issues or inconsistencies based on the Contributor Covenant.

\subsection{Use Cases}
The following use cases illustrate how the tool operates in different situations, ranging from adding a missing code of conduct to updating an outdated version or enhancing an existing one with additional guidelines. 
These use cases demonstrate the tool's ability to assist maintainers in keeping their repositories compliant with ethical standards and ensuring a respectful and inclusive environment for contributors. 
We organize the use cases into two groups based on the underlying service they rely on: 
(1) use cases 1 to 4 rely on \emph{CoCService}, which manages the project's code of conduct, 
and use cases 5 and 6 rely on \emph{ContributionService}, which moderates and analyzes user contributions.

\subsubsection{Use Case 1: Updating an Outdated Code of Conduct}
\begin{figure}[t]
    \centering
    \includegraphics[width=\linewidth]{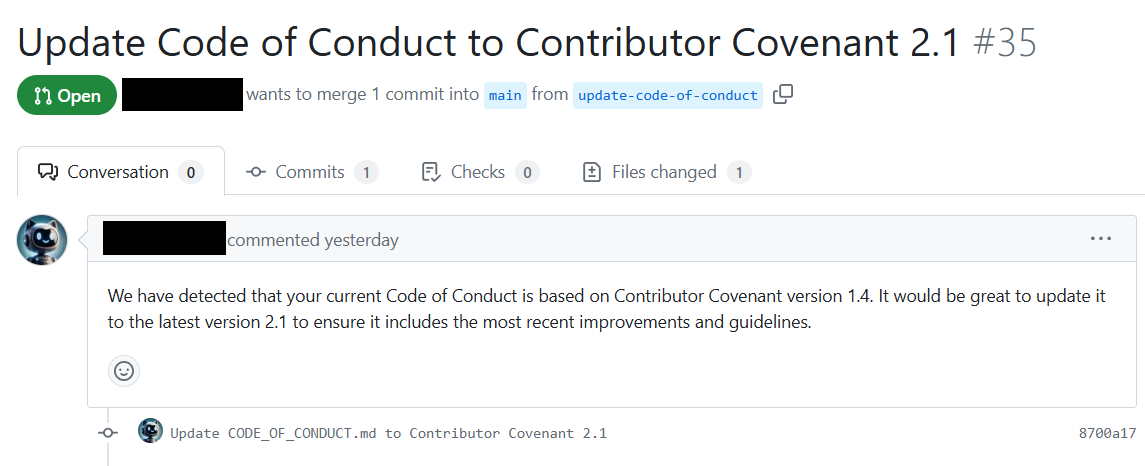}
    \caption{Update CoC to Contributor Covenant 2.1.}
    \label{fig:use_case_update_coc}
\end{figure}

In the scenario where the tool detects an outdated version of the Contributor Covenant (i.e., step \circled{3c} in Figure~\ref{fig:orchestration}), it automatically creates a pull request to update the \texttt{CODE\_OF\_CONDUCT.md} file to the latest version (i.e., version 2.1). 
This ensures that the project follows the most recent ethical guidelines.
As shown in Figure~\ref{fig:use_case_update_coc}, the tool generates a pull request with a message detailing the proposed update. 
This is especially relevant for OSS projects that have adopted older versions of the Contributor Covenant, but may not be actively updating their documentation. 
By automating this process, the tool helps maintainers keep their projects in line with evolving community standards.

\subsubsection{Use Case 2: Adding a Code of Conduct}
\begin{figure}[t]
    \centering
    \includegraphics[width=\linewidth]{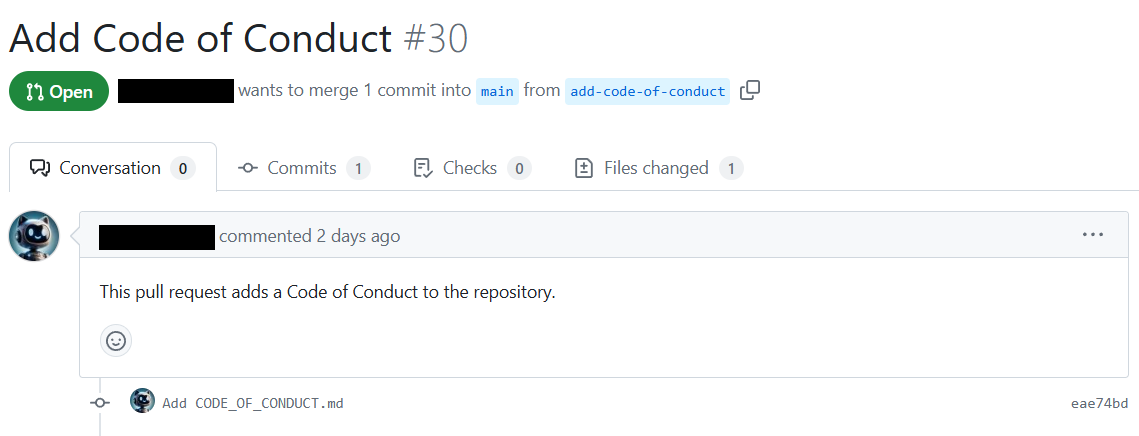}
    \caption{Add Code of Conduct to Repository.}
    \label{fig:use_case_add_coc}
\end{figure}

If no code of conduct is detected in a repository (i.e., step \circled{2b} in Figure~\ref{fig:orchestration}), the tool generates a pull request to add one. 
This pull request introduces the Contributor Covenant as the project's official code of conduct.
Figure~\ref{fig:use_case_add_coc} shows an example of this use case, where the bot automatically creates a branch, adds the code of conduct file, and submits the pull request for the maintainers to review. 
This ensures that every project has a base level of ethical guidelines, promoting a healthy and inclusive environment for contributors.

\subsubsection{Use Case 3: Enhancing an Existing Code of Conduct}
\begin{figure}[t]
    \centering
    \includegraphics[width=\linewidth]{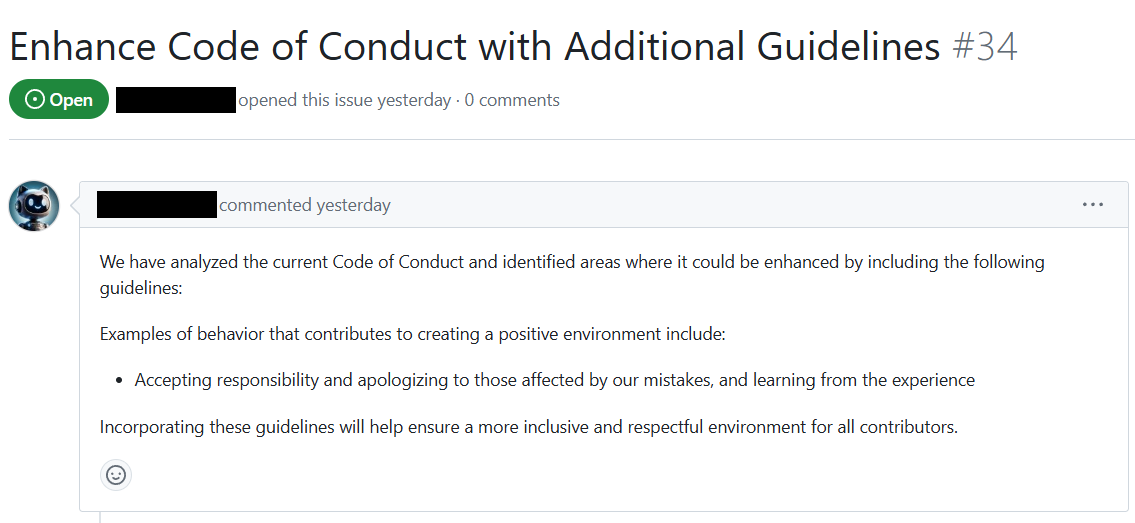}
    \caption{Enhance Code of Conduct with Additional Guidelines.}
    \label{fig:use_case_enhance_coc}
\end{figure}

In cases where a code of conduct is present but missing certain ethical guidelines, the tool analyzes the content of the code of conduct and identifies areas where improvements can be made (i.e., step \circled{4b} in Figure~\ref{fig:orchestration}). 
The bot then creates an issue that suggests adding specific behavioral guidelines that are important for maintaining a respectful and inclusive community.
Figure~\ref{fig:use_case_enhance_coc} illustrates a scenario where the bot detected missing guidelines related to apologizing for mistakes and taking responsibility. 
The tool recommends incorporating these elements to enhance the overall code of conduct, ensuring that the community adheres to best practices.

\subsubsection{Use Case 4: Requesting Help with the Code of Conduct}
\begin{figure}[t]
    \centering
    \includegraphics[width=\linewidth]{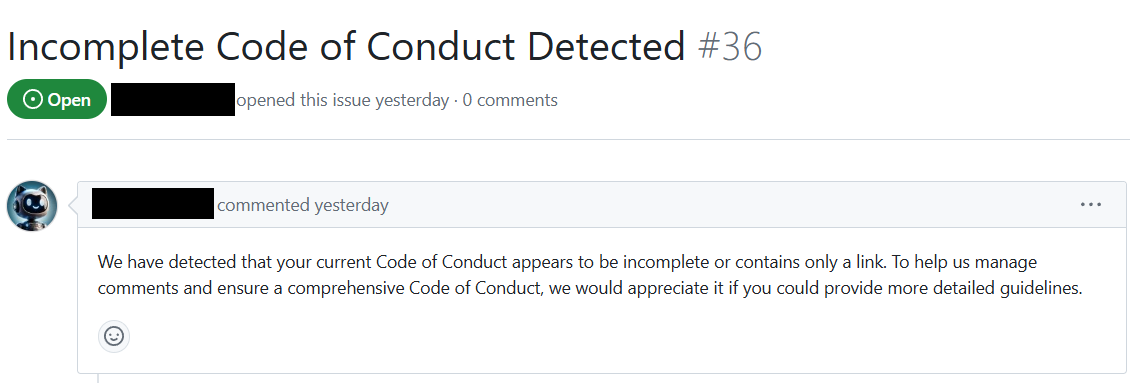}
    \vspace{-1em}
    \caption{Requesting Help with the Code of Conduct.}
    \label{fig:use_case_incomplete_coc}
\end{figure}

The tool handles scenarios where the code of conduct might not be fully analyzable, such as when the code is too short to be considered valid or if it only contains a link to an unavailable external resource. In these cases, the bot creates an issue requesting assistance from the maintainers to provide a more detailed and accessible version of the code of conduct (i.e., step \circled{4c} in Figure~\ref{fig:orchestration}). As shown in Figure~\ref{fig:use_case_incomplete_coc}, the bot prompts the maintainers to update the file or add guidelines directly within the repository to ensure the project's ethical standards are clearly available and reviewable by all contributors.

\subsubsection{Use Case 5: Moderating Inappropriate Contributions}
\rev{When the tool detects a violation of the project's code of conduct within a user contribution (i.e., step \circled{6b} in Figure~\ref{fig:orchestration}), the system automatically identifies the inappropriate content and notifies the project owner.}

\subsubsection{Use Case 6: Providing Positive Feedback}
\begin{figure}[t]
    \centering
    \includegraphics[width=\linewidth]{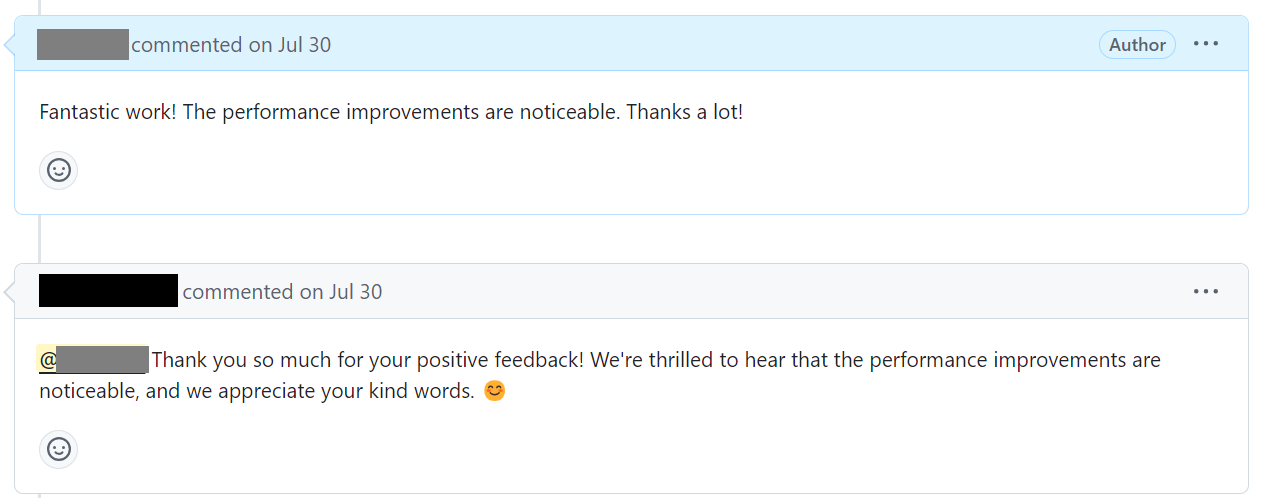}
    \caption{Automated Positive Feedback.}
    \label{fig:use_case_positive_feedback}
\end{figure}

When a contribution complies with the project's code of conduct, the tool generates a response to acknowledge and thank the contributor for their positive engagement (i.e., step \circled{6a} in Figure~\ref{fig:orchestration}). 
Figure~\ref{fig:use_case_positive_feedback} illustrates how the tool generates a thank-you message. 
This not only reinforces positive behavior but also contributes to a welcoming and collaborative environment.

\subsection{\rev{Proof of Concept}}
\rev{We studied the use cases which would be triggered in a real project, such as the \textsc{konya\_ss}\footnote{\url{https://github.com/bmaltais/kohya_ss}} project. As the project does not have a code of conduct, our tool would trigger the use case 2. Regarding the triggering of use cases 5 and 6 (i.e., moderating negative contributions and encouraging posivite ones, respectively), we analyzed the full project and manually extracted a sample of 25 contributions for each flag. Note that not all flags were found (i.e., F6, F8, and F9), while only a limited number of instances were identified for flags F7 (17) and F10 (11). Our tool would have proposed for moderation 66.67\% of the contributions with F7 and 100\% for F10. For positive flags, the tool would have detected on average 43.60\% positive flags, with notable performance for flag F5 (71.43\%). These results highlight room for improvement, related to the experimental nature of our tool. At this early stage, we used the Mixtral-8x22B model to validate the viability of our approach. For a production-level deployment, more meticulous prompt engineering and exploration of alternative models with higher precision, will be necessary.}

%% file: icseseis-related.tex
The role and impact of codes of conduct in OSS communities have been extensively studied, especially regarding their adoption and effectiveness. 
Tourani et al.~\cite{DBLP:conf/wcre/TouraniAS17} were among the first to investigate the presence, scope, and influence of codes of conduct, identifying five common dimensions: purpose, honorable behavior, unacceptable behavior, enforcement, and scope. 
Singh et al.~\cite{DBLP:journals/sqj/SinghBB22} conducted a similar analysis, revealing that less than 10\% of OSS projects adopt a code of conduct, with significant variation in terminology, length, and enforcement. 
\rev{However, these studies mainly focus on the presence and manual enforcement of codes of conduct, while our work introduces an automated bot-based approach based on a set of ethical flags extracted from the Contributor Covenant. Furthermore, our approach not only monitors but also enforces and proactively updates codes of conduct according to well-known recommendations for codes of conduct, improving the ability to maintain healthy collaborative environments.}

Vasilescu et al.~\cite{DBLP:conf/icse/VasilescuFS15} highlighted the importance of team diversity and composition on GitHub, noting the positive impact these factors have on collaboration. 
Li et al.~\cite{DBLP:journals/pacmhci/LiPFD21} expanded on these studies by analyzing how codes of conduct are enforced, showing that manual enforcement by maintainers remains the dominant practice. 
\rev{While our approach shares the concern for community interactions, our solution differs by integrating an automatic approach based on bots that not only detects violations but also encourages continuous updates to codes of conduct, ensuring that projects remain aligned with the latest ethical standards, such as the Contributor Covenant.} 

Other research has explored moderation practices in OSS, such as the work by Hsieh et al.~\cite{Hsieh2023}, who investigated conflict management strategies in OSS projects, which  range from punitive to reformative measures. 
\rev{Our approach focuses on detecting, notifying and correcting inappropriate behavior, but it could be improved by adding moderation and conflict resolution elements inspired by Hsieh et al.'s findings, enhancing our system's ability to handle conflicts more effectively.}

Additionally, studies on user perceptions of automated content moderation provide further insight into how trust and legitimacy can be established in AI-driven systems. 
Pan et al.~\cite{Pan2022} found that users often perceive algorithmic moderation as impartial, but trust is strengthened when human oversight is included. 
Similarly, Molina and Sundar~\cite{Molina2022} emphasized that trust in AI moderation increases when there are mechanisms for user feedback and transparency about decision-making processes. 
Ozanne et al. \cite{Ozanne2022} also noted that while AI is often trusted for clear-cut cases, ambiguity can reduce trust and accountability, highlighting the importance of providing clear explanations for decisions.
\rev{Our current approach is based on an orchestration process which is publicly available, thus promoting external revision, but this approach could guide the development of future moderation capabilities in our system, ensuring that trust and transparency remain central to managing community behavior effectively}.

The work by Shihab et al.~\cite{Shihab2022} provided an overview of bot usage in software engineering, highlighting challenges in bot coordination, human-bot collaboration, and privacy issues. 
\rev{We focused on automating code of conduct management, but our approach could evolve by addressing the challenges identified by Shihab et al., particularly in terms of bot coordination and privacy in collaborative environments}. 
These improvements would allow for better integration of bots into OSS project workflows and strengthen human-bot collaboration.

Our approach to identifying ethical flags in codes of conduct is related to dictionary-based methods used to categorize texts. 
For instance, Loomba et al.~\cite{loomba2023} highlighted the effectiveness of custom dictionaries for evaluating opinions and categorizing themes of interest.
Similarly, Nandwani et al.~\cite{nandwani2021} demonstrated that lexicon-based approaches are particularly useful for extracting thematic categories from unstructured text.
Additionally, Imran et al.~\cite{imran2022} found that lexicographic methods can provide an effective framework for identifying specific indicators in large data volumes

Recent studies on the use of LLMs for sentiment analysis have demonstrated their effectiveness in detecting nuanced emotions, which are critical for analyzing social interactions in OSS communities.
Carneros-Prado et al.~\cite{Carneros2023} conducted a comparative study between GPT-3.5 and IBM Watson, revealing that GPT-3.5 performed competitively and exhibited better adaptability in recognizing complex sentiments without requiring fine-tuning. 
Nadi et al.~\cite{Nadi2024} explored the performance of GPT-3.5 in sentiment analysis of social media data, showing that LLMs can outperform traditional methods while eliminating the need for extensive pre-processing. 
These findings are relevant to the use of LLMs in our approach to analyze OSS contributions, as they provide evidence of their potential for accurate and real-time sentiment analysis.

Regarding prompt engineering, Tran and Matsui~\cite{Tran2024} have highlighted the importance of this technique in enhancing the performance of LLMs for specific tasks, including sentiment analysis. 
Their approach aligns well with our efforts to refine how we prompt LLMs to analyze OSS contributions and identify ethical flags. 
Their study helped us to improve the accuracy and relevance of our sentiment analysis, helping to promote the robustness of our tool.

Finally, Win et al.~\cite{Win2023} introduced Etor, a tool for detecting unethical behaviors in OSS.  
Etor detects ethical violations of a legal nature, while we focus on contributor interactions related to rules in codes of conduct.
Moreover, whereas Win et al.'s approach is reactive, our tool detects violations as they occur and proactively encourages the continuous updating of codes of conduct, fostering a positive and collaborative environment.

%% file: icseseis-conclusion.tex
In this paper, we have presented an approach to provide automatic support for managing codes of conduct in OSS projects.
\rev{We have presented a study on the presence of codes of conduct in OSS projects, how they align with the Contributor Covenant, and a set of ethical flags to drive the management of codes of conduct.}
\rev{Our approach relies on a bot-based solution that helps with the tasks of defining, monitoring, and enforcing codes of conduct}.
We have implemented the approach, where bots are deployed as GitHub actions, and communicate with server-side tools to provide computational resources.
\rev{We believe that our approach also contributes to the broader objective of making software development field more inclusive and ethical.}
 
\rev{As future work, we are interested in improving the precision of the classifier, as identified during the execution of the proof of concept, and exploring mechanisms to customize the orchestration of the approach, thus allowing project maintainers to define enforcement processes for their own codes of conduct.  }
We plan to extend the approach to improve its scalability and to support other social-coding platforms, such as GitLab or Bitbucket; where we will also study the presence of codes of conduct.
The support for additional languages is also being considered, as the current approach only works in English.
\rev{Finally, we plan to evaluate the approach in running OSS projects, to assess its effectiveness and the acceptance by the community (e.g., how it affects to contributors' trust and willingness).}
\rev{Related to this, we are interested in conducting a comparison between bot-based and human-based approaches to manage and enforce codes of conduct}
\rev{This assessment will also help us to identify a potential broader impact of our approach, for instance, whether the use of tools like ours may help to build better OSS commnuities.}

%% file: ms.bbl
\begin{thebibliography}{10}
\providecommand{\url}[1]{#1}
\csname url@samestyle\endcsname
\providecommand{\newblock}{\relax}
\providecommand{\bibinfo}[2]{#2}
\providecommand{\BIBentrySTDinterwordspacing}{\spaceskip=0pt\relax}
\providecommand{\BIBentryALTinterwordstretchfactor}{4}
\providecommand{\BIBentryALTinterwordspacing}{\spaceskip=\fontdimen2\font plus
\BIBentryALTinterwordstretchfactor\fontdimen3\font minus
  \fontdimen4\font\relax}
\providecommand{\BIBforeignlanguage}[2]{{%
\expandafter\ifx\csname l@#1\endcsname\relax
\typeout{** WARNING: IEEEtran.bst: No hyphenation pattern has been}%
\typeout{** loaded for the language `#1'. Using the pattern for}%
\typeout{** the default language instead.}%
\else
\language=\csname l@#1\endcsname
\fi
#2}}
\providecommand{\BIBdecl}{\relax}
\BIBdecl

\bibitem{DBLP:conf/wcre/TouraniAS17}
P.~Tourani, B.~Adams, and A.~Serebrenik, ``Code of conduct in open source
  projects,'' in \emph{Int. Conf. on Software Analysis, Evolution and
  Reengineering}, 2017, pp. 24--33.

\bibitem{DBLP:conf/icse/VasilescuFS15}
B.~Vasilescu, V.~Filkov, and A.~Serebrenik, ``Perceptions of diversity on git
  hub: {A} user survey,'' in \emph{Int. Workshop on Cooperative and Human
  Aspects of Software Engineering}, 2015, pp. 50--56.

\bibitem{Hsieh2023}
J.~Hsieh, J.~Kim, L.~Dabbish, and H.~Zhu, ``Nip it in the bud: Moderation
  strategies in open source software projects and the role of bots,''
  \emph{{ACM} Hum. Comput. Interact.}, vol.~7, no. {CSCW2}, pp. 1--29, 2023.

\bibitem{DBLP:journals/pacmhci/LiPFD21}
R.~Li, P.~Pandurangan, H.~Frluckaj, and L.~Dabbish, ``Code of conduct
  conversations in open source software projects on github,'' \emph{{ACM} Hum.
  Comput. Interact.}, vol.~5, no. {CSCW1}, pp. 19:1--19:31, 2021.

\bibitem{Ehmke2018}
C.~A. Ehmke, ``Codes of conduct now guide open source communities,''
  \url{https://internethealthreport.org/2019/codes-of-conduct-in-open-source-communities}
  (Accessed on 04/Oct/24), 2018.

\bibitem{ContributorCovenant}
C.~A. Ehmke \emph{et~al.}, ``Contributor covenant code of conduct,''
  \url{http://contributor-covenant.org} (Accessed on 04/Oct/24).

\bibitem{DBLP:journals/sqj/SinghBB22}
V.~Singh, B.~Bongiovanni, and W.~Brandon, ``Codes of conduct in open source
  software - for warm and fuzzy feelings or equality in community?''
  \emph{Softw. Qual. J.}, vol.~30, no.~2, pp. 581--620, 2022.

\bibitem{Shihab2022}
E.~Shihab, S.~Wagner, M.~A. Gerosa, M.~Wessel, and J.~Cabot, ``The present and
  future of bots in software engineering,'' \emph{{IEEE} Softw.}, vol.~39,
  no.~5, pp. 28--31, 2022.

\bibitem{Lin2024}
H.~Y. Lin, P.~Thongtanunam, C.~Treude, and W.~Charoenwet, ``Improving automated
  code reviews: Learning from experience,'' in \emph{Int. Conf. on Mining
  Software Repositories}, 2024, pp. 278--283.

\bibitem{Wessel2020}
M.~Wessel, A.~Serebrenik, I.~Wiese, I.~Steinmacher, and M.~A. Gerosa, ``Quality
  gatekeepers: Investigating the effects of code review bots on pull request
  activities,'' \emph{Empir. Softw. Eng.}, vol.~27, no.~5, p. 108, 2022.

\bibitem{ErlenhovNL22}
L.~Erlenhov, F.~G. de~Oliveira~Neto, and P.~Leitner, ``Dependency management
  bots in open-source systems - prevalence and adoption,'' \emph{PeerJ Comput.
  Sci.}, vol.~8, p. e849, 2022.

\bibitem{Damian2024}
D.~Damian, K.~Blincoe, D.~Ford, A.~Serebrenik, and Z.~Masood, \emph{Equity,
  Diversity, and Inclusion in Software Engineering: Best Practices and
  Insights}.\hskip 1em plus 0.5em minus 0.4em\relax Apress Berkeley, 2024.

\bibitem{Albusays2021}
K.~Albusays, P.~Bjorn, L.~Dabbish, D.~Ford, E.~Murphy-Hill, A.~Serebrenik, and
  M.-A. Storey, ``The diversity crisis in software development,'' \emph{{IEEE}
  Software}, vol.~38, no.~2, pp. 19--25, 2021.

\bibitem{McIlwain2019}
C.~D. McIlwain, \emph{Black Software: The Internet and Racial Justice, from the
  AfroNet to Black Lives Matter}.\hskip 1em plus 0.5em minus 0.4em\relax Oxford
  University Press, 2019.

\bibitem{Young2021}
J.-G. Young, A.~Casari, K.~McLaughlin, M.~Z. Trujillo, L.~Hébert-Dufresne, and
  J.~P. Bagrow, ``Which contributions count? analysis of attribution in open
  source,'' in \emph{Int. Conf. on Mining Software Repositories}, 2021, pp.
  242--253.

\bibitem{Casari2021}
A.~Casari, K.~McLaughlin, M.~Z. Trujillo, J.-G. Young, J.~P. Bagrow, and
  L.~Hébert-Dufresne, ``Open source ecosystems need equitable credit across
  contributions,'' \emph{Nat. Comput. Sci.}, vol.~1, no.~2, pp. 67--73, 2021.

\bibitem{HippocraticLicense2021}
C.~A. Ehmke, ``The hippocratic license 3.0: An ethical license for open
  source,'' \url{https://firstdonoharm.dev/} (Accessed on 04/Oct/24), 2021.

\bibitem{Finley2022}
K.~Finley, ``Decisions, decisions: Principles for making important choices in
  open source,'' \url{https://github.com/readme/guides/making-decisions-in-OSS}
  (Accessed on 04/Oct/24), 2022.

\bibitem{Maenpaa2018}
H.~Mäenpää, S.~Mäkinen, T.~Kilamo, T.~Mikkonen, T.~Männistö, and
  P.~Ritala, ``Organizing for openness: six models for developer involvement in
  hybrid oss projects,'' \emph{J. Internet Serv. Appl.}, vol.~9, no.~1, pp.
  17:1--17:14, 2018.

\bibitem{Buritica2019}
J.~P. Buriticá, ``The good, the bad, and the ugly of making decisions in open
  source: Using rfcs to support decision-making when working in public,''
  \url{https://github.com/readme/guides/making-decisions-in-OSS} (Accessed on
  04/Oct/24), 2019.

\bibitem{FariasICSOB2021}
V.~Farias, I.~Wiese, and R.~Santos, ``Power relations within an open source
  software ecosystem,'' in \emph{Int. Conf. on Software Business}, vol. 434,
  2021, pp. 187--193.

\bibitem{nandwani2021}
P.~Nandwani and R.~Verma, ``A review on sentiment analysis and emotion
  detection from text,'' \emph{Soc. Netw. Anal. Min.}, vol.~11, no.~1, p.~81,
  2021.

\bibitem{imran2022}
M.~Imran, A.~Ashraf, F.~Zahra, and A.~Habib, ``A comprehensive survey on
  sentiment analysis: Approaches, challenges and trends,'' \emph{Knowl. Based
  Syst.}, vol. 226, p. 107134, 2021.

\bibitem{Carneros2023}
D.~Carneros{-}Prado, L.~Villa, E.~Johnson, C.~C. Dobrescu, A.~Barrag{\'{a}}n,
  and B.~Garc{\'{\i}}a{-}Mart{\'{\i}}nez, ``Comparative study of large language
  models as emotion and sentiment analysis systems: {A} case-specific analysis
  of {GPT} vs. {IBM} watson,'' in \emph{Int. Conf. on Ubiquitous Computing {\&}
  Ambient Intelligence}, vol. 842, 2023, pp. 229--239.

\bibitem{Nadi2024}
F.~Nadi, H.~Naghavipour, T.~Mehmood, A.~B. Azman, J.~A. Nagantheran, K.~S.~K.
  Ting, N.~M. I. B.~N. Adnan, R.~A. Sivarajan, S.~A. Veerah, and R.~F. Rahmat,
  ``Sentiment analysis using large language models: {A} case study of
  {GPT-3.5},'' in \emph{Int. Conf. on Data Science and Emerging Technologies},
  vol. 191, 2023, pp. 161--168.

\bibitem{Pan2022}
C.~A. Pan, S.~Yakhmi, T.~P. Iyer, E.~Strasnick, A.~X. Zhang, and M.~S.
  Bernstein, ``Comparing the perceived legitimacy of content moderation
  processes: Contractors, algorithms, expert panels, and digital juries,''
  \url{https://doi.org/10.48550/arXiv.2202.06393} (Accessed on 04/Oct/24),
  2022.

\bibitem{Molina2022}
M.~D. Molina and S.~S. Sundar, ``When ai moderates online content: effects of
  human collaboration and interactive transparency on user trust,'' \emph{J.
  Comput. Mediat. Commun.}, vol.~27, no.~4, 2022.

\bibitem{Ozanne2022}
M.~Ozanne, A.~Bhandari, and D.~DiFranzo, ``Shall ai moderators be made visible?
  perception of accountability and trust in moderation systems on social media
  platforms,'' \emph{Big Data Soc.}, vol.~9, no.~2, p. 205395172211156, 2022.

\bibitem{loomba2023}
S.~Loomba, M.~Dave, H.~Arolkar, and S.~Sharma, ``Sentiment analysis using
  dictionary-based lexicon approach: Analysis on the opinion of indian
  community for the topic of cryptocurrency,'' \emph{Annals of Data Science},
  2023.

\bibitem{Tran2024}
V.~Tran and T.~Matsui, ``Improving {LLM} prompting with ensemble of
  instructions: {A} case study on sentiment analysis,'' in \emph{Int. Symp. on
  New Frontiers in Artificial Intelligence}, vol. 14741, 2024, pp. 299--305.

\bibitem{Win2023}
H.~M. Win, H.~Wang, and S.~H. Tan, ``Towards automated detection of unethical
  behavior in open-source software projects,'' in \emph{European Software
  Engineering Conf. and Symp. on the Foundations of Software Engineering},
  2023, pp. 644--653.

\end{thebibliography}
